\documentclass{llncs}
\usepackage{llncsdoc}
\usepackage{array}
\usepackage{colortbl}
\usepackage{cite}
\usepackage[noend]{algpseudocode}
\usepackage[ruled]{algorithm} 
\usepackage{paralist}
\usepackage{amssymb}
\usepackage{amsmath}
\usepackage{etoolbox}
\usepackage{xspace}
\usepackage{microtype}
\usepackage{tikz}
\usepackage{pgfplots}
\usepackage{pgfplotstable}
\usepackage{stmaryrd}
\usepackage{longtable}
\usepackage{listings}
\usepackage{booktabs}
\usepackage{cleveref}
\usepackage[center]{caption}
\usepackage{subcaption}
\usepackage{verbatim}

\crefname{section}{}{}
\Crefname{section}{}{}

\usetikzlibrary{graphs, positioning,shapes,backgrounds}
\usetikzlibrary{intersections}
\usetikzlibrary{calc}

\definecolor{listinggray}{gray}{0.9}
\definecolor{lbcolor}{rgb}{0.9,0.9,0.9}
\definecolor{selected}{rgb}{0.85,0.12,0.12}

\tikzstyle{program_node}=[circle,draw=blue!50,thick,minimum size=6mm]
\tikzstyle{large_program_node}=[circle,draw=blue!50,thick,minimum size=15mm]
\tikzstyle{transition}=[->,thick]
\tikzstyle{selected_transition}=[->,very thick,densely dashed, draw=selected]

\newcommand{\setof}[1]{\ensuremath{\left \{{#1}\right\}}}
\newcommand{\tuple}[1]{\ensuremath{\left( #1 \right) }}

\newcommand{\union}{\cup}

\newcommand{\Real}{\ensuremath{\mathbb{R}}}
\newcommand{\cfa}{\textsc{CFA}\xspace}

\algrenewcommand\Return{\State \algorithmicreturn{} }
\algnewcommand{\LineComment}[1]{\vspace{0.4em}\State \(\triangleright\) #1}
\algnewcommand{\CLineComment}[1]{\State \(\triangleright\) #1}

\makeatletter\@ifclassloaded{llncs}{
\AtBeginEnvironment{definition}{\let\itshape\relax}
}{}\makeatother

\makeatletter
\g@addto@macro\normalsize{%
\abovedisplayskip 5.0\p@ \@plus2\p@ \@minus4\p@
\abovedisplayshortskip \z@ \@plus2\p@
\belowdisplayshortskip 3\p@ \@plus2\p@ \@minus2\p@
\belowdisplayskip \abovedisplayskip
}
\g@addto@macro\small{%
\abovedisplayskip 7.5\p@ \@plus2\p@ \@minus4\p@
\abovedisplayshortskip \z@ \@plus2\p@
\belowdisplayshortskip 3\p@ \@plus2\p@ \@minus2\p@
\belowdisplayskip \abovedisplayskip
}
\renewcommand{\paragraph}{%
  \@startsection{paragraph}{4}%
  {\z@}{0.8ex \@plus 0ex \@minus 1ex}{-1em}%
  {\normalfont\normalsize\bfseries}%
}

\def\@IEEEsectpunct{.\ \,}

\newlength{\sectionspace}
\renewcommand\section{\@startsection{section}{1}{\z@}%
                       {-18\sectionspace \@plus -4\sectionspace \@minus -4\sectionspace}%
                       {12\sectionspace \@plus 4\sectionspace \@minus 4\sectionspace}%
                       {\normalfont\large\bfseries\boldmath
                        \rightskip=\z@ \@plus 8em\pretolerance=10000 }}

\makeatother

\begin{document}

\title{\textcolor{gray}{Proc. VMCAI 2016, (c) Springer} \\ %
    Program Analysis with Local Policy Iteration%
\thanks{The research leading to these results has received funding from the
European Research Council under the European
Union's Seventh Framework Programme (FP/2007-2013) / ERC Grant Agreement
nr.~306595 \mbox{``STATOR''.}}}

\author{Egor George Karpenkov\inst{1} \and David Monniaux\inst{1} \and Philipp
    Wendler\inst{2}}
\institute{Univ. Grenoble Alpes, VERIMAG, F-38000 Grenoble, France\\
CNRS, VERIMAG, F-38000 Grenoble, France
\and University of Passau, Passau, Germany}

\maketitle

\begin{abstract}
We present local policy iteration~(LPI),
a new algorithm for deriving numerical invariants
that combines the precision of max-policy iteration with the flexibility
and scalability of conventional Kleene iterations.
It is defined in the Configurable Program Analysis (CPA)
framework, thus allowing inter-analysis communication.

LPI uses adjustable-block encoding in order to traverse loop-free program sections,
possibly containing branching, without introducing extra abstraction.
Our technique operates over any template linear constraint domain,
including the interval and octagon domains;
templates can also be derived from the program source.

The implementation is evaluated on a set of benchmarks from
the International Competition on Software Verification (SV-COMP).
It competes favorably with state-of-the-art analyzers.
\end{abstract}

\section{Introduction}
\label{sec:motivation}
Program analysis by \emph{abstract interpretation}~\cite{abstract_interpretation}
derives facts about the execution of programs
that are always true regardless of the inputs.
These facts are proved using \emph{inductive invariants},
which satisfy both the initial condition and the transition relation,
and thus always hold.
Such invariants are found within an \emph{abstract domain},
which specifies what properties of the program can be tracked.
Classic abstract domains for numeric properties include [products of] intervals
 and octagons~\cite{DBLP:journals/lisp/Mine06},
both of which are instances of \emph{template linear constraint domains}~\cite{template_constraints_domain}.

Consider classic abstract interpretation with intervals over the program
\texttt{int i=0; while(i < 1000000) i++;}
After the first instruction, the analyzer has a \emph{candidate invariant} $i \in [0,0]$. Going through the
loop body it gets $i \in [1,1]$, thus by least upper bound with the previous
state $[0,0]$ the new candidate invariant is $i \in [0,1]$.
Subsequent \emph{Kleene iterations} yield $[0,2]$, $[0,3]$ etc.
In order to enforce the convergence within a reasonable time, a \emph{widening operator} is used, which extrapolates this sequence to $[0,+\infty)$.
Then, a \emph{narrowing iteration} yields $[0,100000]$.
In this case, the invariant finally obtained is the best possible, but the same
approach yields the suboptimal invariant $[0,+\infty)$ if an unrelated nested loop
is added to the program: \texttt{while(i<100000){while(unknown()){} i++;}}.
This happens because the candidate invariant obtained with widening is its own post-image
under the nested loop, hence narrowing cannot shrink the invariant.

In general, widenings and narrowings are brittle: a
small program change may result in a different
analysis behavior. Their result is \emph{non-monotone}:
a locally more precise invariant at one point may result in a less precise one elsewhere.

\paragraph*{Max-policy iteration} In contrast, max-policy
iteration~\cite{max_strategy_templates} is
guaranteed to compute the least \emph{inductive} invariant in the given abstract
domain.%
\footnote{It does not, however, necessarily output the strongest (potentially
non-inductive) invariant in an abstract domain, which in general entails solving
the halting problem.}
To compute the bound~$h$ of the invariant~$i \leq h$ for the initial example above,
it considers that $h$ must satisfy $h = \max i' \mbox{
    s.t. } (i'=0) \lor (i'=i + 1 \land i < 10000000 \land i \leq h)$
and computes the least inductive solution of this equation by successively considering separate cases:
\begin{compactenum}[(i)]
\item $h = (\max i' \mbox{ s.t. } i'=0) = 0$, which is not inductive, since one can iterate from $i=0$ to $i=1$.
\item $h = \max i' \mbox{ s.t. } i'=i+1 \land i < 1000000 \land i \leq h$,
    which has two
    solutions over $\Real \union \{\infty, -\infty\}$: $h = -\infty$
    (representing unreachable state, discarded)
    and $h = 1000000$, which is finally inductive.
\end{compactenum}

Earlier presentations of policy iteration solve a sequence of global convex
optimization problems whose unknowns are the bounds (here $h$) at every
program location.
Further refinements~\cite{policy_iteration_path_focusing} allowed
restricting abstraction to a cut-set~\cite{cut_set}
of program locations (a set of program
points such that the control-flow graph contains no cycle once these points are
removed), through a combination with \emph{satisfiability modulo theory} (SMT)
solving.
Nevertheless, a global view of the program was
needed, hampering scalability and combinations with other analyses.

\paragraph*{Contribution}
We present the new local-policy-iteration algorithm (LPI) for
computing inductive invariants using policy iteration.
Our implementation is integrated inside the open-source
CPAchecker~\cite{cpachecker} framework for software verification
and uses the maximization-modulo-theory solver $\nu Z$~\cite{opt_z3}.
To the best of our knowledge, this is the first policy-iteration implementation
that is capable of dealing with C code.
We evaluate LPI and show its competitiveness with state-of-the-art analyzers
using benchmarks from the International Competition on Software Verification (SV-COMP).

Our solution improves on earlier max-policy approaches:

\noindent \textbf{(i) Scalability} LPI constructs optimization queries that are at most of
the size of the largest loop in the program.
At every step we only solve the optimization problem necessary for deriving the
\emph{local} candidate invariant.

\noindent \textbf{(ii) Ability to cooperate with other analyses}
LPI is defined within the Configurable Program Analysis (CPA)~\cite{cpa} framework,
which is designed to allow easy inter-analysis collaboration.
Expressing policy iteration as a fixpoint-propagation algorithm
establishes a common ground with other approaches (lazy abstraction, bounded
model checking) and allows communicating with other analyses.

\noindent \textbf{(iii) Precision} LPI uses adjustable-block
encoding~\cite{adjustable_block_encoding},
and thus benefits from the precision offered by SMT solvers,
effectively checking executions of loop-free program segments
without the need for over-approximation.
\emph{Path focusing}~\cite{policy_iteration_path_focusing} has the
same advantage, but at the cost of pre-processing the control-flow graph, which
significantly hinders inter-analysis communication.

\paragraph*{Related Work}
Policy iteration is not as widely used as classic abstract interpretation and (bounded) model checking.
Roux and Garoche~\cite{roux_policy_iteration_integration} addressed
a similar problem of embedding the policy-iteration procedure inside an
abstract interpreter, however their work has a different focus (finding
quadratic invariants on relatively small programs)
and the policy-iteration algorithm remains fundamentally un-altered.
The tool \textsc{ReaVer}~\cite{ReaVer} also performs policy iteration, but focuses on
efficiently dealing with logico-numerical abstract domains; it only
operates on Lustre programs.
The ability to apply policy iteration on strongly connected components one
by one was (briefly) mentioned before~\cite{pi_separate_scc}.
Our paper takes the approach significantly further, as our value-determination problem is
more succinct, we apply the principle of
locality to the policy-improvement phase, and we formulate policy iteration as
a classic fixpoint-iteration algorithm, enabling communication with other
analyses.
Finally, it is possible to express the search for an inductive invariant
as a nonlinear constraint solving problem~\cite{direct_invariant_generation} or
as a quantifier elimination problem~\cite{Monniaux_LMCS10}, but both these
approaches scale poorly.

\section{Background}
\label{sec:background}
We represent a program $P$ as a control flow automaton (\cfa)
$\tuple{\mathit{nodes}, X, \mathit{edges}}$, where $\mathit{nodes}$ is a set of control
states, and $X = \setof{x_1,\dots,x_n}$ are the variables of~$P$.
Each edge $e \in \mathit{edges}$  is a tuple $\tuple{A, \tau(X, X'), B}$, where
$A$ and $B$ are nodes,
and $\tau(X, X')$ is a \emph{transition relation}: a formula
defining the semantics of a transition over the set of input variables $X$ and fresh output variables $X'$.
A \emph{concrete state} of the program $P$ is a map $X \to \mathbb{Q}$ from
variables to rationals\footnote{%
We support integers as well, as explained in Sec.~\ref{sec:implementation_details}.}.
A set~$C$ of concrete states is represented using a first-order formula $\phi$ with
free variables from $X$, such that for all $c \in C$ we have $c \models \phi$.

\paragraph*{Template Linear Constraint Domains}
A \emph{template linear constraint} is a linear inequality $t \cdot X \leq b$
where $t$ is a vector of constants (\emph{template}), and $b$ is an unknown.
A \emph{template linear constraint domain}~\cite{template_constraints_domain}
(TCD) is
an abstract domain defined by a matrix of coefficients $a_{ij}$, which
determines what template linear constraints are expressible within the domain:
each row~$t$ of the matrix is a template (the word ``template'' also refers to the symbolic product $t \cdot X$, e.g. $i + 2j$).
An abstract state in a TCD is defined by a vector
$(d_1,\dots,d_m)$ and represents the set
$\bigwedge_{i=1}^m t_i \cdot X \leq d_i$ of concrete states.
The $d_i$'s range over extended rationals ($\Real
\union \{\infty, -\infty\}$), where positive infinity represents unbounded templates and
negative infinity represents unreachable abstract states.
The domain of \emph{products of intervals} is one instance of TCD, where
the templates are $\pm x_i \leq c_i$ for program variables $x_i$.
The domain of \emph{octagons}~\cite{DBLP:journals/lisp/Mine06} is another, with
templates $\pm x_i \pm x_j$ and $\pm x_i$.
Any template linear constraint domain is a subset of the domain of convex polyhedra~\cite{cousot78}.

The strongest abstract postcondition in a TCD is
defined by optimization: maximizing all templates subject to the
constraints introduced by the
semantics of the transition and the previous abstract state.
For the edge $e = \tuple{A, \tau(X, X'), B}$, previous abstract state~$D = \tuple{d_1, \dots, d_m}$, and a set~$\setof{t_1, \dots, t_m}$ of templates,
the output abstract state is $D' = \tuple{d'_1, \dots, d'_m}$ with
\[d'_i = (\max t_i \cdot X' \mbox{ s.t. } \textstyle\bigwedge\nolimits_i t_i \cdot X \leq d_i \land \tau(X, X'))\]

For example, for the abstract state $i \leq 0 \land j \leq 0$ under the transition
$i' = i + 1 \land i \leq 10$ the new abstract state is $i \leq d^i \land y \leq d^j$, where
$d^i = \max i' \mbox{ s.t. } i \leq 0 \land j \leq 0 \land i' = i + 1 \land i <
10 \land j' = j$
and
$d^j$ is the result of maximizing $j'$ subject to the same constraints.
This gets simplified to $i \leq 1 \land j \leq 0$.

Kleene iterations in a TCD (known as \emph{value iterations}) may
fail to converge in finite time, thus the use of \emph{widenings}, which result in hard-to-control imprecision.

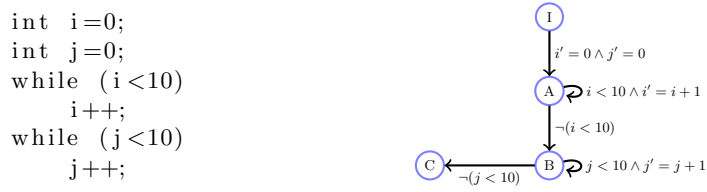
\begin{figure}[t]
    \centering
    \begin{minipage}{0.35\linewidth}
        \begin{lstlisting}{c}
int i=0;
int j=0;
while (i<10)
    i++;
while (j<10)
    j++;
        \end{lstlisting}
    \end{minipage}
    \begin{minipage}{0.5\linewidth}
        \centering
        \begin{tikzpicture}[auto, scale=0.6, transform shape]
    \node (i) [program_node] {I};
    \node (a) [program_node, below=1 of i] {A};
    \node (b) [program_node, below=1 of a] {B};
    \node (c) [program_node, left=2 of b] {C};

    \path [transition] (i) edge node {$i' = 0 \land j' = 0$} (a)
        (a) edge [loop right] node  {$i < 10 \land i' = i + 1$} (a)
        (a) edge node {$\lnot (i < 10)$} (b)
        (b) edge [loop right] node {$j < 10 \land j' = j + 1$} (b);
    \path [transition] (b) edge node {$\lnot (j < 10)$} (c);
\end{tikzpicture}
    \end{minipage}
    \vspace{-2mm}
    \caption{Running example -- C program and the corresponding \cfa}
    \label{fig:running_example}
\end{figure}

\paragraph*{Policy Iteration} Policy iteration addresses the convergence problem
of \emph{value-iteration} algorithms by operating on an equation system that an
inductive invariant has to satisfy.
Consider the running example shown in Fig.~\ref{fig:running_example}.
Suppose we analyze this program with the
templates $\setof{i, j}$, and look for the least inductive invariant
$D = (d^i_A, d^j_A, d^i_B, d^j_B)$ that satisfies the following for all
possible
executions of the program ($x_N$ denotes the value of the variable~$x$ at the node~$N$):
\begin{equation*}
i_A \leq d^i_A \land i_B \leq d^i_B \land j_A \leq d^j_A \land j_B \leq d^j_B
\end{equation*}
To find it, we solve for the smallest $D$
that satisfies the \emph{fixpoint equation [system]}
for the running example,
stating that the set of abstract states represented by $D$ is equal to its strongest
postcondition within the abstract domain:
\[
\begin{split}
    d^{i}_{A} =& \sup{i'} \mbox{ s. t. } (i' = 0 \land j'=0) \\
        \lor & (i \leq d^{i}_{A} \land j \leq d^j_A \land i < 10 \land i'=i
                  + 1 \land j'=j) \lor \bot \\
    d^{j}_{A} =& \sup{j'} \mbox{ s. t. } (i' = 0 \land j'=0) \\
        \lor &
        (i \leq d^{i}_{A} \land j \leq d^j_A \land i < 10 \land i' = i
                  + 1 \land j' = j) \lor \bot \\
    d^{i}_{B} =& \sup{i'} \mbox{ s. t. } (\lnot (i < 10) \land i \leq d^i_A
    \land j' \leq d^j_A \land i'=i) \\
          \lor &
               (i \leq d^i_B \land j \leq d^j_B \land j < 10 \land j' = j +
               1 \land i' = i) \lor \bot \\
    d^{j}_{B} =& \sup{j'} \mbox{ s. t. } (\lnot (i < 10) \land i \leq d^i_A
    \land j' \leq d^j_A \land i'=i) \\
    \lor &
               (i \leq d^i_B \land j \leq d^j_B \land j < 10 \land j' = j +
               1 \land i'=i) \lor \bot
\end{split}
\]
Note the equation structure:
\begin{inparaenum}[(i)]
    \item Disjunctions represent non-deterministic choice for a new value.
    \item The argument $\bot$ is added to all disjunctions, representing
        infeasible choice, corresponding to the bound value $-\infty$.
    \item Supremum is taken because the bound must be higher than all the
        possible options, and it has to be~$-\infty$ if no choice is
        feasible.
\end{inparaenum}

A simplified equation system with each disjunction replaced by 
one of its arguments is called a \emph{policy}.
The least solution of the whole equation system is the least solution of at
least one policy
(obtained by taking the solution, and picking one argument for each disjunction,
such that the solution remains unchanged).
Policy iteration finds the least tuple of unknowns ($d$'s)
satisfying the fixpoint equation by iterating over possible policies, and
finding a solution for each one.

For program semantics consisting of linear assignments and possibly
non-deterministic guards
it is possible to find a fixpoint of each policy using one linear programming
step.
This is based on the result that for a  monotone and concave function%
\footnote{Order-concave in the presence of multiple templates,
    see~\cite{max_strategy_templates} for detailed discussion}
$f$ and $x_0$ such that $f(x_0) > x_0$, the least fixpoint of $f$
greater than $x_0$ can be computed in a single step%
\footnote{Over rationals, we discuss the applicability to integers in
    Sec.~\ref{sec:implementation_details}}.

It is possible to solve the global equation system by solving all
(exponentially many) policies one by one.
Instead, \emph{policy iteration}~\cite{max_strategy_templates} computes solutions
for a sequence of policies;
each solution is guaranteed to be less than the least solution of the original equation system,
and the solutions form an ascending sequence.
The iteration starts with the policy having least possible value
($\bot$ for each disjunction,
the solution is $-\infty$ assignment to all unknowns), and eventually terminates
when a solution of the original equation system (an inductive invariant) is found.
The termination is guaranteed as there is only a finite number of solutions.

For each policy the algorithm finds a \emph{global value}: the least fixpoint in the template
constraints domain of the reduced equation system.
For instance, in the running example, for the policy
$d^i_A = \sup{i'} \mbox{ s. t. } i' = 0 \land j' = 0$ (only one unknown is
shown for brevity) the global value is $d^i_A = 0$.
This step is called \emph{value determination}.
After the global value is computed the algorithm checks whether the policy can be
\emph{improved}: that is, whether we can find another policy that will yield a
larger value \emph{than the previously obtained global value}.
In the running example we want to test the following policy
for the possibility of improvement:
\[d^i_A = \sup{i'} \mbox{ s. t. } (i \leq d^{i}_{A} \land j \leq d^i_A \land i <
10 \land i' = i + 1 \land j' = j)\]
We do so by computing the \emph{local value}: substituting the unknown ($d^i_A$) on the
right hand side with the value from the previously obtained global value, and
checking whether the result is greater than the previously obtained bound.
In our example we get the local value $d^i_A = 1$, which is indeed an
improvement over $d^i_A = 0$ (\emph{policy-improvement} step).
After the policy is selected, we go back to the value-determination step, obtaining
$d^i_A = 10$, and we repeat the process until convergence (reaching a step where no
policy can be further improved).

Under the
assumption that the operations on the edges can be expressed as conjunctions of
linear (in)equalities,
it can be shown~\cite{max_strategy_templates} that:
\begin{inparaenum}[(i)]
\item The value-determination step can be performed with linear programming.
\item The resulting value is an under-approximation of the least inductive
    invariant.
\item Each policy is selected at most once and
    the final fixed point yields the least inductive invariant in the domain.
\end{inparaenum}

\begin{example}[Policy-Iteration Trace on the Running Example]
We solve for the unknowns $(d^i_A, d^j_A, d^i_B, d^j_B)$, defining a (global) abstract value~$v$.

In our example, disjunctions arise from multiple
incoming edges to a single node, hence a policy is defined by a choice of an incoming
edge per node per template, or $\bot$ if no such choice is feasible.
We represent a policy symbolically as a 4-tuple of predecessor nodes (or $\bot$),
as there are two nodes, with two policies to be chosen per node.
The order corresponds to the order of the tuple of the unknowns.
The initial policy $s$ is $(\bot, \bot, \bot, \bot)$.
The trace on the example is:
    \begin{compactenum}
        \item Policy improvement: $s = (I, I, \bot, \bot)$, obtained with a
            local value $(0, 0, -\infty, -\infty)$.
        \item Value determination: corresponds to the initial condition 
            $v = (0, 0, -\infty, -\infty)$.
        \item Policy improvement: $s = (A, I, \bot, \bot)$, selecting the
            looping edge, local value is $(1, 0, -\infty, -\infty)$.
        \item Value determination: accelerates the loop convergence to
            $v = (10, 0, -\infty, -\infty)$.
        \item Policy improvement: $s = (A, I, A, A)$, with a local value $(10,
            0, 10, 0)$ finally there is a
            feasible policy for the templates associated with the node $B$.
        \item Value determination: does not affect the result $v = (10,
            0, 10, 0)$.
        \item Policy improvement: select the second looping edge: $s = (A, I, A,
            B)$ obtaining a local value $(10, 0, 10, 1)$.
        \item Value determination: accelerate the second loop to $v = (10, 0, 10,
            10)$.
        \item Finally, the policy cannot be improved any further and we terminate.
    \end{compactenum}

 On this example we could have obtained the
 same result by Kleene iteration,
 but in general the latter might fail to converge within finite time.
 The usual workaround is to use heuristic widening,
 with possible and hard-to-control imprecision.
Our value-determination step can be seen as a widening that provides an under-approximation to the least fixed point.

    Each policy improvement requires at least four (small) linear programming
    (LP) queries, and each value determination requires one (rather large)
    LP query.
    \label{ex:running-example-trace}
\end{example}

\paragraph*{Path Focusing and Large-Block Encoding}
In traditional abstract interpretation and policy iteration, the obtained
invariant is expressed as an abstract state at each \cfa node.
This can lead to a significant loss in precision: for instance, since most
abstract domains only express convex properties, it is impossible to express
$|x| \geq 1$, which is necessary to prove this assertion:
\texttt{if (abs(x) >= 1) { assert(x != 0); }}

This loss can be recovered by reducing the number of ``intermediate''
abstract states by allowing more expressive formulas associated with edges.
Formally, two consecutive edges $(A, \tau_1(X, X'), B)$ and $(B, \tau_2(X, X'),
C)$, with no other edges incoming or outgoing to $B$ can be merged into one edge
 $(A, \tau_1(X, \hat{X}) \land \tau_2(\hat{X}, X'), C)$.
Similarly, two parallel edges $(A, \tau_1(X, X'), B)$ and $(A, \tau_2(X,X'),
B)$, with no other edges incoming to $B$ can be replaced by a new edge
$(A, \tau_1(X, X') \lor \tau_2(X, X'), B)$.
For a well-structured \cfa, repeating this transformation in a fixpoint manner
(until no more edges can be merged) will lead to a new \cfa where the only
remaining nodes are loop heads.

Such a transformation was shown to increase both precision and performance for
model checking~\cite{large_block_encoding}.
Adjustable block encoding~\cite{adjustable_block_encoding} gets the same
advantages without the need for \cfa pre-processing.
Independently, the approach was applied with the same result to Kleene
iterations~\cite{path_focusing} and to max-policy
iterations~\cite{policy_iteration_path_focusing}.
In fact, the \cfa in Fig.~\ref{fig:running_example} was already
reduced in this manner for the ease of demonstration.

On the reduced \cfa the number of possible policies associated with a single
edge becomes exponential,
and explicitly iterating over them is no longer feasible.
Instead, the path focusing approach uses a \emph{satisfiability modulo theory}
(SMT) solver to select an improved policy.

\paragraph*{Configurable Program Analysis}
\textsc{CPA}~\cite{cpa} is a framework for expressing algorithms performing
program analysis.
It uses a generic fixpoint-computation algorithm,
which is \emph{configured} by a given analysis.
We formulate LPI as a \textsc{CPA}.

The \textsc{CPA} framework makes no assumptions on the performed analysis,
thus many analyses were successfully expressed and implemented
within it, such as bounded model checking, abstract
interpretation and k-induction (note that an analysis defined within the framework is
also referred to as a \textsc{CPA}).

Each \textsc{CPA} configures the fixpoint algorithm by providing
an \emph{initial abstract state}, a \emph{transfer relation}
(specifying how to produce successors),
a \emph{merge operator} (specifying whether and how to merge abstract states),
and a \emph{stop operator} (specifying whether a newly produced abstract state is
covered).
The algorithm keeps a set of reached abstract states and a list of ``frontier''
abstract states,
and at each step produces successor states from the frontier states using the \emph{transfer relation},
and then tries to merge the new states with existing states using the
\emph{merge operator}.
If a new state is covered by the set of reached states
according to the \emph{stop operator},
it is discarded, otherwise it is added to the set of reached states
and the list of frontier states.
We show the CPA algorithm as Alg.~\ref{alg:cpa}.

\begin{algorithm}[t]
    \centering
    \begin{algorithmic}[1]
        \State \textbf{Input}: a CPA $\tuple{D, \mbox{transfer-relation},
                \mbox{merge}, \mbox{stop}}$, an initial abstract state $e_0 \in
            E$
            (let $E$ denote the set of elements of the semi-lattice of $D$)
        \State \textbf{Output}: a set of reachable abstract states
        \State \textbf{Variables}: a set reached of elements of $E$, a set
        $\mbox{waitlist}$ of elements of $E$
        \State $\mbox{waitlist} \gets \setof{e_0}$
        \State $\mbox{reached} \gets \setof{e_0}$
        \While{$\mbox{waitlist} \neq \emptyset$}
            \State Pop $e$ from $\mbox{waitlist}$
            \ForAll{$e' \in \mbox{transfer-relation}(e)$}
                \ForAll{$e'' \in \mbox{reached}$}
                    \LineComment{Combine with existing abstract state}
                    \State $e_\text{new} \gets \mbox{merge}(e', e'')$
                    \If{$e_\text{new} \neq e''$}
                        \State $\mbox{waitlist} \gets (\mbox{waitlist} \union
                        \setof{e_\text{new}}) \setminus \setof{e''}$
                        \State $\mbox{reached} \gets (\mbox{reached} \union
                        \setof{e_\text{new}}) \setminus \setof{e''}$
                    \EndIf
                \EndFor
                \LineComment{Whether $e'$ is already covered by existing states}
                \If{$\lnot \mbox{stop}(e', \mbox{reached})$}
                    \State $\mbox{waitlist} \gets \mbox{waitlist} \union
                    \setof{e'}$
                    \State $\mbox{reached} \gets \mbox{reached} \union
                    \setof{e'}$
                \EndIf
            \EndFor
        \EndWhile
        \Return $\mbox{reached}$
    \end{algorithmic}
    \caption{CPA Algorithm (taken from~\cite{cpa})}
    \label{alg:cpa}
\end{algorithm}

\section{Local Policy Iteration (LPI)}
\label{sec:local_policy_iteration}

The running example presented in the background
(Ex.~\ref{ex:running-example-trace})
has four value-determination steps and five policy-improvement steps.
Each policy-improvement step corresponds to at most \#policies
$\times$ \#templates $\times$ \#nodes LP queries, and each
value-determination step requires solving an LP problem with at least
\#policies $\times$ \#templates $\times$ \#nodes variables.
Most of these queries are redundant, as the updates propagate only
\emph{locally} through the \cfa: there is no need to re-compute the policy
if no new information is available.

We develop a new policy-iteration-based algorithm, based on the principle of
\emph{locality},
which aims to address the scalability issues and the problem of communicating
invariants with other analyses.
We call it \emph{local policy iteration} or \textsc{LPI}.
To make it scalable, we consider the structure of a \cfa being analyzed, and we
aim to exploit its \emph{sparsity}.

A large majority of (non-recursive) programs are well-structured:
they consist of statements and possibly nested loops.
Consider checking a program $P$ against an error property $E$.  If $P$ has no
loops, it can be converted into a single formula~$\Psi(X')$, and an SMT solver can be
queried for the satisfiability of $\Psi(X') \land E(X')$, obtaining either
a counter-example or a proof of unreachability of $E$.
However, in the presence of loops, representing all concrete states reachable by
a program as a
formula over concrete states in a decidable first-order logic
is impossible, and abstraction is required.
For example, bounded model checkers unroll the loop, lazy-abstraction-based
approaches partially unroll the loop and use the predicates from Craig interpolants to ``cover'' future unrollings, and
abstract interpretation relies on abstraction within an abstract domain.

In LPI, we use the value-determination step to
``close'' the loop and compute the fixpoint value for the given policy.
Multiple iterations through the loop might be necessary to find the optimal
policy and reach the global fixpoint.
In the presence of nested loops, the process is repeated in a fixpoint manner:
we ``close'' the inner loop, ``close'' the outer loop with the new information
from the inner loop available, and repeat the process until convergence.
Each iteration selects a new policy,
thus the number of possible iterations is bounded.

Formally, we state LPI as a Configurable Program Analysis (\textsc{CPA}),
which requires defining the lattice of abstract
states, the transfer relation, the merge operator, and the stop operator.
The \textsc{CPA} for LPI is intended to be used in combination with other CPAs
such as a \textsc{CPA} for tracking location information (the program counter),
and thus does not need to keep track of this information itself.
To avoid losing precision, we do not express the invariant
as an abstract state at every node: instead the transfer
relation operates on formulas and we only perform over-approximation at certain
\emph{abstraction points} (which correspond to loop heads in a well-structured
\textsc{\cfa}).
This approach is inspired by adjustable-block
encoding~\cite{adjustable_block_encoding}, which performs the same operation for
predicate abstraction.
One difference to path focusing~\cite{path_focusing} is that we still traverse
intermediate nodes, which facilitates inter-analysis communication.

We introduce two lattices: \emph{abstracted states} (not to be confused
    with \emph{abstract states} in general: both intermediate and abstracted
    states are \emph{abstract}) for states
associated with abstraction points (which can only express abstract states in the template
constraints domain)
and \emph{intermediate states} for all others
(which can express arbitrary concrete state spaces using decidable SMT formulas).

An \emph{abstracted state} is an element of a template
constraints domain with meta-information added to record the \emph{policy} being used.

\begin{definition}[Abstracted State]
    An abstracted state is a mapping from the externally given set~$T$ of templates
    to tuples $\tuple{d, \mathsf{policy},
    \mathsf{backpointer}}$, where $d \in \Real$ is a bound for the
    associated template~$t$ (the represented property is $t\cdot X \leq d$),
    $\mathsf{policy}$ is a formula representing the policy that was used for deriving $d$
    ($\mathsf{policy}$ has to be monotone and concave, and in particular contain
    no disjunctions), and $\mathsf{backpointer}$ is an abstracted state that
    is a starting point for the $\mathsf{policy}$
    (base case is an empty mapping).
    \label{def:abstracted_state}
\end{definition}

The preorder on abstracted states is defined by component-wise
comparison
of bounds associated with respective templates (lack of a bound
corresponds to an unbounded template).
The concretization is given by the
conjunction of represented template linear constraints, disregarding
$\mathsf{policy}$ and $\mathsf{backpointer}$ meta-information.
For example, an abstracted state $\{ x: \tuple{10, \_, \_} \}$ (underscores
represent meta-information irrelevant to the example) concretizes to
\mbox{$\{ c \mid c[x] \leq 10 \}$},
and the initial abstracted state $\{ \}$ concretizes to all concrete states.

Intermediate states represent reachable state-spaces using formulas directly,
again with meta-information added to record the ``used'' policy.

\begin{definition}[Intermediate State]
    An \emph{intermediate state} is a tuple $\tuple{a_0, \phi}$, where $a_0$ is
    a \emph{starting} abstracted state, and $\phi(X, X')$
    is a formula over a set of input variables $X$
    and output variables $X'$.
    \label{def:intermediate_state}
\end{definition}

The preorder on intermediate states is defined by syntactic comparison only:
states with identical starting states and identical formulas are deemed equal,
and incomparable otherwise.
The concretization is given by satisfiable assignments to $X'$
subject to $\phi(X,X')$
and the constraints derived from $a_0$ applied to input variables $X$.
For example, an intermediate state $\tuple{\{ x: \tuple{10, \_, \_} \}, x'=x+1}$
concretizes to the set $\{ c \mid c[x] \leq 11 \}$ of concrete states.

\emph{Abstraction} (Alg.~\ref{alg:abstraction})
is the conversion of an intermediate state~$\tuple{a_0, \phi}$ to
an abstracted state, by maximizing all templates~$t\in T$
subject to constraints introduced by $a_0$ and $\phi$,
and obtaining a backpointer and a policy from the produced model~$\mathcal{M}$.
This amounts to selecting the appropriate disjuncts in each disjunction of~$\phi$.
To do so, we annotate $\phi$ with \emph{marking variables}:
each disjunction $\tau_1 \lor \tau_2$ in~$\phi$ is replaced by $(m \land \tau_1)
\lor (\lnot m \land \tau_2)$ where $m$ is a fresh propositional variable.
A policy associated to a bound is then identified by the
values of the marking variables at the optimum (subject to the constraints
introduced by $\phi$ and $a_0$),
and is obtained by replacing the marking variables in $\phi$ with their values
from~$\mathcal{M}$.
Thus the abstraction operation effectively performs the \emph{policy-improvement}
operation for the given node, as only the policies which are
feasible with respect to the current candidate invariant (given by previous
abstracted state) are selected.

\begin{example}[LPI Propagation and Abstraction]
    Let us start with an abstracted state $a = \{x: \tuple{100, \_,
        \_}\}$ (which
    concretizes to $\{ c \mid c[x] \leq 100 \}$, underscores stand for some
    policy and some starting abstracted state) and a set~$\setof{x}$ of templates.

    After traversing a section of code
    \texttt{if(x <= 10){x += 1;} else {x = 0;}}
    we get an intermediate state $\tuple{a, \phi}$ with
    $\phi = (x \leq 10 \land x'=x+1 \lor x > 10 \land x'=0)$ and a backpointer
    to the starting \emph{abstracted state} $a$.
    Suppose in our example the given C code fragment ends with a loop head.
    Then we use \emph{abstraction} (Alg.~\ref{alg:abstraction}) to convert the intermediate state
    $\tuple{a, \phi}$ into a new abstracted state.

    Firstly, we annotate $\phi$ with \emph{marking variables}, which are used to
    identify the selected policy, obtaining $x \leq 10 \land x'=x+1 \land m_1 \lor x > 10
    \land x'=0 \land \lnot m_1$.
    Afterwards, we optimize the obtained formula (together with the constraints
    from the starting abstracted state $a$) for the highest values of templates.
    This amounts to a single OPT-SMT query: \[\sup{x'} \mbox{ s.t. } x \leq 100
    \land (x \leq 10 \land x'=x+1 \land m_1 \lor x > 10
    \land x'=0 \land \lnot m_1)\]

    The query is satisfiable with a maximum of $11$, and an SMT model
    $\mathcal{M}:$
    \mbox{$\{x':11, m_1: \mbox{true}, x:10\}$}.
    Replacing the marking variable~$m_1$ in~$\phi$ with its value
    in~$\mathcal{M}$ gives us a
    disjunction-free formula $x \leq 10 \land x'=x+1$,
    which we store as a \emph{policy}.
    Finally, the newly created abstracted state is $\{x: \tuple{11, x \leq 10
        \land x'=x+1, a }\}$.
\end{example}

\begin{algorithm}[t]
\begin{algorithmic}[1]
\State \textbf{Input: } intermediate state $\tuple{a_0, \phi}$, set $T$ of templates
\State \textbf{Output: } generated abstracted state $\mathit{new}$
\State $\mathit{new} \gets$ empty abstracted state
\ForAll{$\mbox{template } t \in T$}
    \State $\hat{\phi} \gets \phi$ with disjunctions annotated using a set of
    marking variables $M$
    \LineComment{Maximize subject to the constraints introduced by the formula}
    \CLineComment{and the starting abstracted state.}
    \State $d \gets \max{t \cdot X'}$ subject to $\hat{\phi}(X,X') \land a_0$
    \State $\mathcal{M} \gets$ model at the optimal
    \LineComment{Replace marking variables $M$ in $\hat{\phi}$ with their value from the model $\mathcal{M}$,}
    \CLineComment{generating a concave formula that represents the policy.}
    \State Policy $\psi \gets \hat{\phi}[M/\mathcal{M}]$
    \State $\mathit{new}[t] \gets \tuple{d, \psi, a_0}$
\EndFor
\Return $\mathit{new}$
\end{algorithmic}
\caption{LPI Abstraction}
\label{alg:abstraction}
\end{algorithm}

The local \emph{value-determination} step (Alg.~\ref{alg:value_determination})
computes the least fixpoint for the chosen policy across the entire strongly
connected component where the current node $n$ lies.
The algorithm starts with a map \emph{influencing} from nodes to  abstracted
states, which is generated by transitively following policy backpointers,
and converting the resulting set of abstracted states to a map\footnote{The are no collisions as abstracted
    states are joined at nodes.}.
From this map, we generate a global optimization problem, where the set of
fresh variables $d_{n_i}^{t}$ represents the maximal value a template $t$ can obtain
at the node $n_i$ using the policies selected.
Variable $d_{n_i}^t$ is made equal to the namespaced\footnote{%
Namespacing means creating fresh copies by attaching a certain \emph{prefix}
to variable names.} \emph{output} value of the
policy~$\psi(X, X')$ chosen for $t$ at $n_i$
(line~\ref{alg:value_determination:output_constraint}).
For each policy $\psi$ and the associated backpointer~$a_0$, we \emph{constrain}
the \emph{input} variables of $\psi$ using a set of variables $d_{n_0}^{t_0}$
representing bounds at the node~$n_0$ associated with~$a_0$
(line~\ref{alg:value_determination:input_constraint}).
This set of ``input constraints'' for value determination results in a quadratic
number of constraints in terms of the number of selected policies.
Finally, for each template~$t$ we maximize for~$d_n^t$
(line~\ref{alg:value_determination:maximization}),
which is the maximum possible value for~$t$ at node~$n$
under the current policy, and we
record the bound in the generated abstracted state
(line~\ref{alg:value_determination:record}),
keeping the old policy and backpointer.

The local-value-determination algorithm is almost identical to max-strategy
evaluation~\cite{policy_iteration_path_focusing},
except for two changes: we only add potentially relevant constraints from the
``closed'' loop (found by traversing backpointers associated with policies),
and we maximize objectives one by one,
not for their sum (which avoids special casing
infinities, and enables optimizations outlined in
Sec.~\ref{sec:implementation_details}).
Unlike classic policy iteration, we only run local value determination
after merges on loop heads,
because in other cases the value obtained by abstraction is the same as the
value which could be obtained by value determination.

\begin{algorithm}[t]
\begin{algorithmic}[1]
\State \textbf{Input: } node $n$, map $\mathit{influencing}$ from nodes to abstracted states,
set $T$ of templates
\State \textbf{Output: } generated abstracted state $\mathit{new}$
\State $\mathit{constraints} \gets \emptyset$

\ForAll{$\mbox{node } n_i \in \mathit{influencing}$}
    \State $\mbox{state } s \gets \mathit{influencing}[n_i]$
    \ForAll{$\mbox{template } t \in s$}
        \State $\tuple{\mbox{bound } d, \mbox{policy } \psi, \mbox{backpointer } a_0} \gets s[t]$
        \State Generate a unique string $\mathit{namespace}$

        \LineComment{Prefix all variables in $\psi$.}
        \CLineComment{$X'_{\mathit{namespace}}, X_{\mathit{namespace}}$ is a set of namespaced output/input variables for $\psi$.}
        \State $\mathit{constraints} \gets \mathit{constraints} \union \setof{
            \psi[X/X_{\mathit{namespace}}][X'/X'_{\mathit{namespace}}]
            }$
        \State $d_{n_i}^{t} \gets$ fresh variable (upper bound on $t$ at $n$)
        \State $\mathit{constraints} \gets \mathit{constraints} \union
            \setof{d_{n_i}^{t} = t \cdot X'_{\mathit{namespace}}}$
                \label{alg:value_determination:output_constraint}
        \State $n_0 \gets$ location associated with $a_0$
        \ForAll{$t_0 \in a_0$}
            \State $\mathit{constraints} \gets \mathit{constraints} \union
                \setof{t_0 \cdot X_{\mathit{namespace}} \leq d_{n_0}^{t_0}}$
                \label{alg:value_determination:input_constraint}
        \EndFor
    \EndFor
\EndFor

\State $\mathit{new} \gets$ empty abstracted state

\ForAll{templates $t \in T$}
    \State $\tuple{d_0, \psi, a_0} \gets \mathit{influencing}[n]$
    \State $d \gets \max d_{n}^t$ subject to $\mathit{constraints}$
        \label{alg:value_determination:maximization}
    \State $\mathit{new}[t] \gets \tuple{d, \psi, a_0}$
        \label{alg:value_determination:record}
\EndFor

\Return $\mathit{new}$
\end{algorithmic}
\caption{Local Value Determination}
\label{alg:value_determination}
\end{algorithm}

\paragraph*{Formulation as a CPA}
The \emph{initial state} is the abstracted state~$\{\}$ (empty map),
representing $\top$ of the template constraints domain.
The \emph{stop operator} checks whether a newly created abstracted state
is covered by one of the existing abstracted states using the preorder described above.
The \emph{transfer relation} finds the successor state for a given \cfa edge.
It operates only on intermediate states -- an abstracted
state $a_0$ is firstly converted to the intermediate state $\tuple{a_0, \mathit{true}}$.
Then, the transfer-relation operator runs symbolic execution:
the successor of an intermediate state $\tuple{a, \phi(X, X')}$ under the edge
$\tuple{A, \tau(X, X'), B}$
is the intermediate state $\tuple{a, \phi'(X, X')}$ with $\phi'(X, X') \equiv \exists
\hat{X} . \phi(X, \hat{X}) \land \tau(\hat{X}, X')$.
If the successor node is a loop head, then \emph{abstraction}
(Alg.~\ref{alg:abstraction}) is performed on the resulting state.

The \emph{merge operator} has two operation
modes, depending on whether we are dealing with abstracted states or with
intermediate states.

For two abstracted states, we perform the join: for each template,
we pick the largest bound out of the two possible,
and we keep the corresponding policy and the backpointer.
If the merge ``closes'' the loop (that is, we merge at the loop head, and one of
the updated policies has a backpointer to a state inside the loop), we find the
map \emph{influencing} by recursively following the backpointers of the joined
state, and run \emph{local value determination}
(Alg.~\ref{alg:value_determination}).
For two intermediate states $\tuple{a_1, \phi_1}$ and $\tuple{a_2, \phi_2}$
with $a_1$ identical to $a_2$
the merge operator returns the disjunction $\tuple{a_1, \phi_1 \lor \phi_2}$.
Otherwise, we keep the states separate.

The local-value-determination problem only contains the constraints resulting
from policies of the abstracted states associated with nodes in the current loop.
This optimization does not affect the invariant as only the nodes
dominating the loop head can change it.
Of those, only the invariants of the nodes reachable from the
loop head can be affected by the computation:
i.e., the strongly connected component of~$n$.

\paragraph*{Properties of LPI}~

\noindent\textbf{Soundness} LPI, like any configurable program analysis,
        terminates when no more updates can be performed, and
        newly produced abstract states are subsumed
        (in the preorder defined by the lattice) by the already discovered ones.
        Thus, it is an inductive invariant: the produced abstract states satisfy the
        initial condition and all successor states are subsumed by the existing
        invariant.
        Hence the obtained invariant is sound.

\noindent\textbf{Termination} An infinite sequence of produced \emph{abstract} states must contain
        infinitely many \emph{abstracted} states, as they are associated with
        loop heads.
        However, each subsequent abstraction on the same node must
        choose a different policy to obtain a successively higher value, but the
        number of policies is finite.
        An infinite sequence is thus impossible, hence termination.

\noindent\textbf{Optimality}
    In the absence of integers, LPI terminates with the
    same invariant as classical policy iteration with
    SMT~\cite{policy_iteration_path_focusing}.
    The outline of the proof is that LPI can be seen as an efficient oracle for
    selecting the next policy to update (note that policies selected by LPI are
    always \emph{feasible} with respect to the current invariant candidate).
    Skipping value-determination steps when they have no effect, and attempting to
    include only relevant constraints in the value-determination problem do not
    alter the values of obtained fixed points.

\begin{example}[LPI Trace on the Running Example]
    We revisit the running example (Fig.~\ref{fig:running_example}) with LPI:
    \begin{compactenum}
        \item We start with the empty abstracted state $a_0 \equiv \{\}$.
        \item Transfer relation under the edge $(I, \phi_1, A)$ produces the new
            intermediate state $\tuple{a_0, i' = 0 \land j' = 0}$ associated with $A$.
            As $A$ is a loop head, we perform an abstraction to obtain the
            abstracted
            state~$a_1 \equiv \{ i: \tuple{0, \_, a_0}, j:  \tuple{0, \_, a_0}\}$
            (corresponding to $i \leq 0 \land j \leq 0$) [2 linear programming problems].
        \item Transfer relation explores the edge $(A, \phi_2, A)$ and produces
            the intermediate state $\tuple{a_1, i \leq 0 \land j' \leq 0 \land i' = i + 1}$.
            Again we perform an abstraction, obtaining the abstracted
            state~$a_2 \equiv \{ i: \tuple{1, \_, a_1}, j: \tuple{0, \_, a_1} \}$
             [2 LP problems].
        \item The merge operator on node $A$ merges the new state $a_2$
            with the previous state $a_1$, yielding the abstracted
            state~$a_3 \equiv \{ i: \tuple{1, \_, a_1}, j: \tuple{0, \_, a_0} \}$.
            Value determination ``closes'' the loop, producing
            $a_4 \equiv \{ i: \tuple{10, \_, a_1}, j: \tuple{0, \_, a_0} \}$.
            \newline [1 LP problem].
        \item Transfer relation explores the edge $(A, \phi_3, B)$ and produces
            the intermediate state $\tuple{a_3, i' \leq 10 \land (\lnot i' < 10) \land j' \leq 0}$,
            which is abstracted
            to \newline $a_5 \equiv \{ i: \tuple{10, \_, a_4}, j: \tuple{0, \_, a_4} \}$
            [2 LP problems].
        \item The edge $(B, \phi_4, B)$ is explored, resulting in the
            intermediate state\\ $\tuple{a_4, i' \leq 10 \land j \leq 0 \land j'=j+1}$, which is
            abstracted
            into \newline $a_6 \equiv \{ i: \tuple{10, \_, a_5}, j: \tuple{1, \_, a_5} \}$
            [2 LP problems].
        \item Value determination produces the state
            $a_7 \equiv \{ i: \tuple{10, \_, a_4}, j: \tuple{10, \_, a_5} \}$,
            and the exploration concludes. [1 LP problem].
    \end{compactenum}

    Compared to the original algorithm there are two value-determination
    problems instead of four, both on considerably smaller scale.
    There are also only ten LP problems, compared to more than
    twenty in the original version.
    The improvement in performance is more than a fixed constant: if the number
    of independent loops in the running example was to increase from $2$ to $N$, the increase
    in the analysis time of classic policy iteration would be quadratic,
    while LPI would scale linearly.
\end{example}

\section{Extensions and Implementation Aspects}
\label{sec:implementation_details}

\paragraph*{Template Synthesis}
The template constraints domain requires templates defined for the given program.
In LPI, we can simulate the interval and octagon domains
by synthesizing templates of the form $\pm x$, $\pm x \pm y$ for
every numeric variable $x, y$ in the program \emph{alive} at the given program node.
Moreover, the templates can be synthesized from error properties: e.g.
for \texttt{assert(x >= 2 * y)} we could generate the templates $\pm (x - 2y)$.

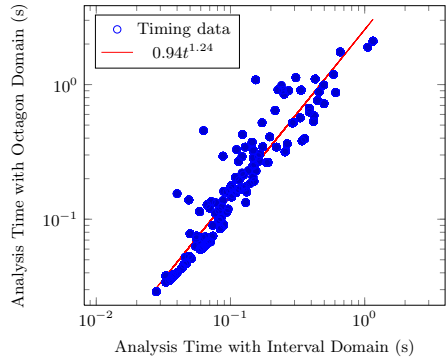
\begin{figure}[t]
    \centering
\begin{tikzpicture}[scale=0.7]
    \begin{axis}[xmode=log, ymode=log, xlabel={Analysis Time with Interval Domain (s)},
            enlarge x limits = 0.05,
            enlarge y limits = 0.05,
    ylabel={Analysis Time with Octagon Domain (s)}, axis equal=true, legend style={legend pos={north west}}]
        
\addplot+[only marks, mark=o, xlabel=Size, ylabel=Time, opacity=1] table {
    X Y
    0.056 0.075
0.342 0.382
0.050 0.078
0.427 1.100
0.147 0.248
0.186 0.263
0.220 0.343
0.174 0.345
0.386 0.665
0.251 0.847
0.165 0.304
0.039 0.039
0.115 0.269
1.154 2.103
0.265 0.363
0.148 0.323
0.102 0.178
0.103 0.171
0.110 0.158
0.114 0.167
0.255 0.316
0.035 0.035
0.033 0.038
0.060 0.065
0.059 0.114
0.033 0.034
0.088 0.294
0.038 0.039
0.331 0.566
0.164 0.271
0.337 0.910
0.045 0.046
0.056 0.072
0.417 0.532
0.049 0.139
0.657 1.747
0.457 0.880
0.047 0.047
0.047 0.048
0.190 0.264
0.290 0.519
0.148 0.286
0.609 0.873
0.067 0.128
0.085 0.086
0.094 0.159
0.101 0.176
0.497 0.722
0.098 0.166
0.116 0.166
0.086 0.124
0.258 0.319
0.074 0.090
0.065 0.074
0.294 0.523
0.049 0.053
0.145 0.373
0.094 0.113
0.038 0.038
0.445 0.761
0.129 0.343
0.334 0.907
0.043 0.043
0.044 0.044
0.051 0.051
0.084 0.086
0.074 0.075
0.065 0.064
0.070 0.121
0.083 0.097
0.062 0.063
0.046 0.052
0.102 0.146
0.084 0.085
0.055 0.063
0.060 0.060
0.028 0.029
0.040 0.040
0.037 0.037
0.078 0.132
0.061 0.061
0.084 0.085
0.073 0.074
0.089 0.091
0.071 0.068
0.072 0.135
0.082 0.099
0.130 0.174
0.063 0.062
0.062 0.073
0.124 0.170
0.037 0.038
0.084 0.086
0.085 0.102
0.061 0.060
0.036 0.036
0.068 0.067
0.059 0.060
0.077 0.111
0.068 0.069
0.082 0.116
0.215 0.642
0.151 0.254
0.154 1.084
0.388 0.624
0.358 0.397
0.109 0.205
0.196 0.409
0.125 0.213
1.055 1.895
0.161 0.293
0.307 1.127
0.271 0.901
0.495 0.991
0.586 1.191
0.419 0.588
0.226 0.915
0.155 0.228
0.122 0.292
0.130 0.133
0.123 0.426
0.096 0.119
0.116 0.195
0.172 0.521
0.131 0.161
0.245 0.908
0.103 0.152
0.147 0.206
0.089 0.161
0.132 0.159
0.111 0.329
0.040 0.155
0.063 0.455
0.150 0.191
0.066 0.066
0.069 0.067
0.060 0.060
0.140 0.184
0.088 0.138
0.136 0.230
0.122 0.221
0.240 0.988
};

\addplot+[mark= ] table[y=create col/linear regression={y=Y}] {
    X Y
    0.056 0.075
0.342 0.382
0.050 0.078
0.427 1.100
0.147 0.248
0.186 0.263
0.220 0.343
0.174 0.345
0.386 0.665
0.251 0.847
0.165 0.304
0.039 0.039
0.115 0.269
1.154 2.103
0.265 0.363
0.148 0.323
0.102 0.178
0.103 0.171
0.110 0.158
0.114 0.167
0.255 0.316
0.035 0.035
0.033 0.038
0.060 0.065
0.059 0.114
0.033 0.034
0.088 0.294
0.038 0.039
0.331 0.566
0.164 0.271
0.337 0.910
0.045 0.046
0.056 0.072
0.417 0.532
0.049 0.139
0.657 1.747
0.457 0.880
0.047 0.047
0.047 0.048
0.190 0.264
0.290 0.519
0.148 0.286
0.609 0.873
0.067 0.128
0.085 0.086
0.094 0.159
0.101 0.176
0.497 0.722
0.098 0.166
0.116 0.166
0.086 0.124
0.258 0.319
0.074 0.090
0.065 0.074
0.294 0.523
0.049 0.053
0.145 0.373
0.094 0.113
0.038 0.038
0.445 0.761
0.129 0.343
0.334 0.907
0.043 0.043
0.044 0.044
0.051 0.051
0.084 0.086
0.074 0.075
0.065 0.064
0.070 0.121
0.083 0.097
0.062 0.063
0.046 0.052
0.102 0.146
0.084 0.085
0.055 0.063
0.060 0.060
0.028 0.029
0.040 0.040
0.037 0.037
0.078 0.132
0.061 0.061
0.084 0.085
0.073 0.074
0.089 0.091
0.071 0.068
0.072 0.135
0.082 0.099
0.130 0.174
0.063 0.062
0.062 0.073
0.124 0.170
0.037 0.038
0.084 0.086
0.085 0.102
0.061 0.060
0.036 0.036
0.068 0.067
0.059 0.060
0.077 0.111
0.068 0.069
0.082 0.116
0.215 0.642
0.151 0.254
0.154 1.084
0.388 0.624
0.358 0.397
0.109 0.205
0.196 0.409
0.125 0.213
1.055 1.895
0.161 0.293
0.307 1.127
0.271 0.901
0.495 0.991
0.586 1.191
0.419 0.588
0.226 0.915
0.155 0.228
0.122 0.292
0.130 0.133
0.123 0.426
0.096 0.119
0.116 0.195
0.172 0.521
0.131 0.161
0.245 0.908
0.103 0.152
0.147 0.206
0.089 0.161
0.132 0.159
0.111 0.329
0.040 0.155
0.063 0.455
0.150 0.191
0.066 0.066
0.069 0.067
0.060 0.060
0.140 0.184
0.088 0.138
0.136 0.230
0.122 0.221
0.240 0.988
};

\addlegendentry{Timing data}
\addlegendentry{%
    $\pgfmathprintnumber{\pgfplotstableregressionb}
    t^{\pgfmathprintnumber{\pgfplotstableregressiona}}$}

    \end{axis}
\end{tikzpicture}
    \vspace{-2mm}
    \caption{Octagon vs. Interval LPI Analysis Time
        (Dataset and Setup as in Sec.~\ref{sec:results})}
    \label{fig:oct_vs_intervals}
\end{figure}

We show the analysis time of LPI
(excluding startup and parsing)
in the interval-domain-mode vs.
octagon-domain-mode in Fig.~\ref{fig:oct_vs_intervals}
(each data point corresponds to an analyzed program).
The number of octagon templates is quadratic in terms of the
number of interval templates, thus we expect a quadratic rise in analysis
time, however in practice we observe a sub-quadratic increase.

This has motivated us to experiment with simulating a more expressive domain.
We generate templates $\pm 2x \pm
y$, $\pm x \pm y \pm z$, and even $\pm 2x \pm y \pm z$, for every possible
combination of live variables $x, y, z$ at the given program location.
Using this new ``rich'' template generation strategy we achieve a significant precision
improvement as shown by the number of verified programs in the legend of Fig.~\ref{fig:feature_comparison}.

\paragraph*{Dealing With Integers}
Original publications on max-policy iteration in template constraints domain
deal exclusively with reals,
whereas C programs operate primarily on integers\footnote{%
Previous work~\cite{max_strategy_intervals_integers} deals with finding the exact
interval invariants for programs involving integers, but only for a very
restricted program semantics.}.
Excessively naive handling of integers leads to poor results:
with an initial condition $x=0$,
$x \in [0, 4]$ is inductive for the transition system
$x' = x + 1 \land x \neq 4$ in integers, but not
in rationals, due to the possibility of the transition $x=3.5$ to $x=4.5$.
An obvious workaround is to rewrite each strict inequality $a < b$ into $a \leq
b-1$: on this example, the transition becomes $x = x+1 \land (x \leq 3 \lor x
\geq 5)$ and $x \in [0, 4]$ becomes inductive on rationals.
However, to make use of data produced by an additional \emph{congruence} analysis,
we use optimization modulo theory with integer and real
variables  for abstraction, and mixed integer linear
programming for value determination.

Unfortunately, linear relations over the integers are not \emph{concave}, which is
a requirement for the least fixpoint property of policy iteration.
Thus the encoding described above may still result in an over-approximation.
Consider the following program:
\begin{small}
\begin{lstlisting}{c}
x=0; x_new=unknown();
while (2 * x_new == x+2) {
    x = x_new; x_new = unknown();
}
\end{lstlisting}
\end{small}
LPI terminates with a fixpoint $x \leq 2$, yet the least fixpoint is $x \leq 1$.

\paragraph*{Congruence}
A congruence analysis which tracks
whether a variable is even or odd can be run in parallel with LPI
(a more general congruence analysis may be used, but we did not find the need for it on our examples).
During the LPI abstraction step, the congruence information is conjoined to the formula being maximized,
and the bounds from LPI are used for the congruence analysis.

This combination enhances the precision on our dataset
(cf. Fig.~\ref{fig:feature_comparison}), and demonstrates the usefulness of
expressing policy iteration as a typical fixpoint computation.
Furthermore, it provides a strong motivation to use integer formulas for
integer variables in programs, and not their rational relaxation.

\paragraph*{Optimizations}
In~Sec.~\ref{sec:local_policy_iteration} we describe the local value-determination
algorithm which adds a quadratic number of constraints in terms of policies.
In practice this is often prohibitively expensive.
The quadratic blow-up results from the ``input'' constraints to each policy,
which determine the bounds on the input variables.
We propose multiple optimization heuristics which increase the performance.

As a motivation example, consider a long trace ending with an
assignment \texttt{x = 1}.
If this trace is feasible and chosen as a policy for the template $x$, the
output bound will be $1$, regardless of the input.
With that example in mind, consider the abstraction procedure from which we
derive the bound $d$ for the template~$t$.
Let $\tuple{\_, \phi(X,X')}$ be the intermediate state used for the abstraction
(Alg.~\ref{alg:abstraction}).
We check the satisfiability of $\phi(X,X') \land t \cdot X' > d$;
if the result is unsatisfiable, then the bound of $t$ is
\emph{input-independent}, that is, it is always $d$ if the
trace is feasible.
Thus we do not add the \emph{input constraints} for
the associated policy in the value-determination stage.
Also, when computing the map \emph{influencing} from nodes to abstracted states for the
value-determination problem, we do not follow the backpointers for
input-independent policies, potentially drastically shrinking
the resulting constraint set.
Similarly, if none of the variables of the ``input template'' occur in the
policy, the initial constraint is irrelevant and can be dropped.

Furthermore,
we limit the size of the value-determination LP by merging some of the unknowns.
This is equivalent to equating these variables, thus strengthening the constraints.
The result thus under-approximates the fixed point of the selected policy.
If it is less than the policy fixed point (not inductive with respect to the policy),
we fall back to the normal value determination.

During \emph{abstraction} on the intermediate state $\tuple{a_0, \psi}$,
we may skip the optimization query based on a syntactic check:
if we are optimizing for the template~$t$, and
none of the variables of $t$ occur in~$\psi$, we return the bound
associated with~$a_0[t]$.

Additionally, during maximization we add a redundant lemma to the set of constraints that
specifies that the resultant value has to be strictly larger than the current
bound.
This significantly speeds up the maximization by shrinking the search space.

\paragraph*{Iteration Order} In our experiments, we have found performance to
depend on the iteration order.
Experimentally, we have determined a good iteration order to be the
recursive iteration strategy using the weak topological
ordering~\cite{wto}.
This is a strength of LPI: it blends into existing iteration strategies.

\paragraph*{Unrolling} We unroll loops up to depth $2$, as some invariants can
only be expressed in the template constraints domain in the presence of
unrollings (e.g., invariants involving a variable whose initial value
is set only inside the loop).

\paragraph*{Abstraction Refinement for LPI}
As a template constraints domain can be configured by the number of templates
present, it is a perfect candidate for refinement, as templates can be added to
increase the precision of the analysis.

However, a full abstraction-refinement algorithm for LPI would be outside of the
scope of this work, and thus to obtain the results we use a naive algorithm
that iteratively tries progressively more precise and costly configurations
until the program can be verified.
The configurations we try are (in that order): \begin{inparaenum}[(i)]
\item Intervals
\item Octagons
\item Previous + Unrolling
\item Previous + Rich Templates $(\pm x \pm y \pm z)$
\item Previous + Congruence Analysis.
\end{inparaenum}

\section{Experiments}
\label{sec:results}

We have evaluated our tool on the benchmarks from the category ``Loops''
of the International Competition on Software Verification (SV-COMP'15)~\cite{svcomp15}
consisting of $142$ C programs, $93$ of which are correct (the error property is
unreachable).
We have chosen this category for evaluation because
its programs contain numerical assertions about variables modified in loops,
whereas other categories of SV-COMP
mostly involve variables with a small finite set of possible values
that can be enumerated effectively.
All experiments were performed
with the same resources as in SV-COMP'15:
an Intel Core i7-4770 quad-core CPU with 3.40\,GHz,
and limits of 15\,GB RAM and 900\,s CPU time per program.
The tool is integrated inside the open-source verification framework
CPAchecker~\cite{cpachecker},
used configuration and detailed experimental results are available at
\texttt{http://lpi.metaworld.me}.

We compare LPI (with abstraction refinement) with three tools
representing different approaches to program analysis:
    \textbf{BLAST 2.7.3 (SV-COMP'15)}~\cite{blast_svcomp}, which uses lazy abstraction,
    \textbf{PAGAI (git hash \texttt{254c2fc693})}~\cite{pagai}, which uses abstract interpretation with path
        focusing, and
    \textbf{CPAchecker 1.3.10-svcomp15 (SV-COMP'15)}~\cite{cpachecker}, the winner of SV-COMP
    2015 category ``Overall'',
        which uses an ensemble of different techniques: explicit value, k-induction, and
        lazy predicate abstraction.
For LPI we use CPAchecker in version~1.4.10-lpi-vmcai16.

Because LPI is an incomplete approach, it can only produce safety proofs
(no counter-examples).
Thus in Table~\ref{tab:summary_table} we present the
statistics on the number of safety proofs produced by different tools.
The first five columns represent \emph{differences} between
approaches: the cell corresponding to the row A and a column B (read ``A vs.
B'') displays the number of programs A could verify and B could not.
In the column \emph{Unique} we show the number of programs only the given tool
could verify (out of the analyzers included in the comparison).
The column \emph{Verified} shows the total number of programs a tool could
verify.
The column \emph{Incorrect} shows false positives: programs that contained a
bug, yet were deemed correct by the tool --- our current implementation unsoundly ignores integer overflows, as though the program used mathematical integers.%
\footnote{%
It is possible to add sound overflow handling, as done in e.g. Astr\'ee, to our
approach, at the expense of extra engineering.}

From this table we see that LPI verifies more examples than other tools can, including seven programs that others cannot.

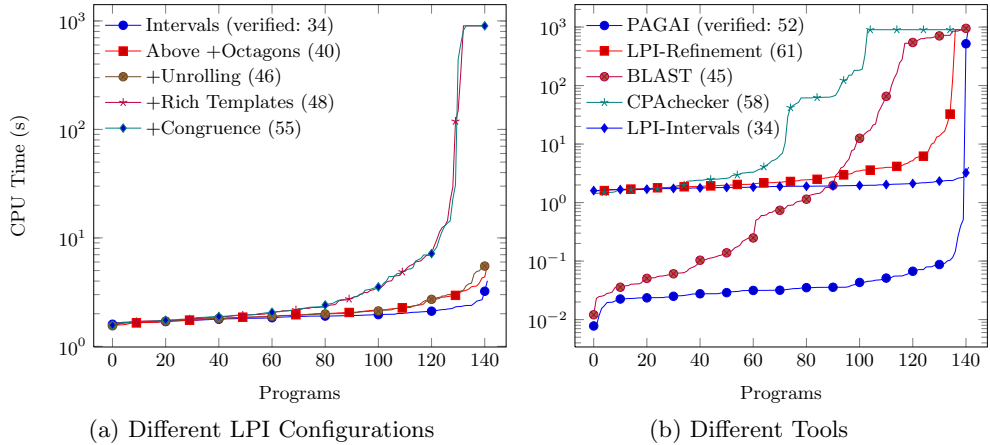
\begin{figure}[t]
    \hspace{-3.3em}\begin{subfigure}[t]{0.6\textwidth}
        \centering
        \begin{tikzpicture}[scale=0.8]
    \pgfplotstableread{results/comparison_across_configs_sorted.dat}\comparisontable
    \begin{axis}[xmode=normal, axis equal=false,  ymode=log,
xlabel={Programs},
ylabel={CPU Time (s)},
enlarge x limits = 0.05,
enlarge y limits = 0.07,
legend entries={
Intervals (verified: 34),
Above +Octagons (40),
+Unrolling (46),
+Rich Templates (48),
+Congruence (55)},
legend style={
    draw=none,
    fill=none,
	cells={anchor=west},
	legend pos={north west},
	font={\footnotesize},
	at={(0.01, 0.99)},
}
]

        \addplot+[mark repeat=20] table[x index=0, y index=1] {\comparisontable};
         \addplot+[mark repeat=20, mark phase=10] table[x index=0, y index=2] {\comparisontable};
       \addplot+[mark repeat=20, mark phase=0] table[x index=0, y index=3] {\comparisontable};
       \addplot+[mark repeat=20, purple, mark phase=10] table[x index=0, y index=4] {\comparisontable};
       \addplot+[mark repeat=20, teal] table[x index=0, y index=5] {\comparisontable};

    \end{axis}
\end{tikzpicture}
        \vspace{-1mm}
        \caption{Different LPI Configurations}
        \label{fig:feature_comparison}
    \end{subfigure}
    \hspace{-1em}\begin{subfigure}[t]{0.5\textwidth}
        \centering
            \begin{tikzpicture}[scale=0.8]
        \pgfplotstableread{results/total_timing_comparison.dat}\comparisontable
        \begin{axis}[xmode=normal, axis equal=false,  ymode=log,
            xlabel={Programs},
            enlarge x limits = 0.05,
            enlarge y limits = 0.07,
            legend entries={
                PAGAI (verified: 52),
                LPI-Refinement (61),
                BLAST (45),
                CPAchecker (58),
                LPI-Intervals (34)
            },
            legend style={
                draw=none,
                fill=none,
                cells={anchor=west},
                legend pos={north west},
                font={\footnotesize},
                at={(0.01, 0.99)},
            }
            ]
            \addplot+[mark repeat=10] table[x index=0, y index=1] {\comparisontable};
            \addplot+[mark repeat=10, mark phase=5] table[x index=0, y index=2] {\comparisontable};
            \addplot+[mark repeat=10, purple, mark phase=0] table[x index=0, y index=3] {\comparisontable};
            \addplot+[mark repeat=10, teal, mark phase=5] table[x index=0, y index=4] {\comparisontable};
            \addplot+[mark repeat=10, mark phase=0] table[x index=0, y index=5] {\comparisontable};
        \end{axis}
\end{tikzpicture}
        \vspace{-1.2em}
        \vspace{-1mm}
        \caption{Different Tools}
        \label{fig:timing_comparison}
    \end{subfigure}
    \vspace{-3mm}
    \caption{Quantile Timing Plots.\\Each data point is an analyzed program, timeouts are excluded.}
\end{figure}

\paragraph*{Timing Results}
In~Sec.~\ref{sec:implementation_details} we have described the various possible configurations of LPI.
As trying all possible combinations of features is exponential,
tested configurations represent cumulative stacking of features.
We present the timing comparison across those in the quantile plot in
Fig.~\ref{fig:feature_comparison}, and in the legend we report the number of programs each
configuration could verify.
Each data point is an analyzed
program, and the series are sorted separately for each configuration.

\begin{table}[t]
    \centering
    
\definecolor{LightGray}{gray}{0.9}
\begin{tabular}{l|rrrr|rrr}
   & vs. PAGAI & LPI & BLAST & CPAchecker & Unique & \textbf{Verified} &
   \textbf{Incorrect} \\
\midrule
\textbf{PAGAI} & \cellcolor{LightGray}  & 4 & 13 & 15 & 1 & 52 & 1\\
\textbf{LPI} & 13 & \cellcolor{LightGray}  & 20 & 20 & 7 & 61 & 1\\
\textbf{BLAST} & 6 & 4 & \cellcolor{LightGray}  & 8 & 0 & 45 & 1\\
\textbf{CPAchecker} & 21 & 17 & 21 & \cellcolor{LightGray}  & 12 & 58 & 2\\

\midrule
\end{tabular}

    \caption{Number of verified programs of different tools\protect\linebreak(LPI in abstraction-refinement mode)}
    \label{tab:summary_table}
    \vspace{-5mm}
\end{table}

The quantile plot for timing comparison across different tools is shown in
Fig.~\ref{fig:timing_comparison}.
We have included two LPI configurations in the comparison:
fastest (LPI-Intervals) and the most precise one (LPI-Refinement, switches to a
more expensive strategy out of the ones in
Fig.~\ref{fig:feature_comparison} if the program cannot be verified).
From the plot we can see that LPI performance compares favorably with lazy
abstraction, but that it is considerably outperformed by abstract
interpretation.
The initial difference in the analysis time between the
\textsc{CPAchecker}-based tools and others is due to \textsc{JVM} start-up time
of about $2$ seconds.

\section{Conclusion and Future Work}
We have demonstrated that LPI is a viable approach to
program analysis, which can outperform state-of-the-art competitors either
in precision (abstract interpretation), or both in precision and scalability
(predicate abstraction).
However, much work needs to be done to bring policy-iteration-based approaches
to the level of maturity required for analyzing industrial-scale codebases, in
particular:
\begin{compactitem}
    \item Sound handling of machine integers and floats, and overflow checking
        in particular.
        The only incorrect result given by LPI on the dataset was due to the
        unsound overflow handling.
        It is possible to check the obtained invariants for inductiveness
        using bitvectors or overflow checks.
    \item Template abstract domains are perfect candidates for \emph{refinement}:
        dynamically adding templates during the analysis.
        Using counter-examples and refining the domain using
        \textsc{CEGAR}~\cite{CEGAR}
        approach is a promising research direction.
\end{compactitem}

\paragraph*{Acknowledgments} The authors wish to thank Tim King
for proof-reading and extremely valuable feedback, Nikolaj
Bj{\o}rner for improving $\nu Z$ performance on our difficult cases,
and the anonymous reviewers for their helpful suggestions.

\bibliographystyle{ieeetr}
\bibliography{library}

\clearpage

\end{document}